\documentclass[lettersize,journal]{IEEEtran}
\usepackage{amsmath,amsfonts}
\usepackage{algorithmic}
\usepackage{algorithm}
\usepackage{array}
\usepackage{textcomp}
\usepackage{stfloats}
\usepackage{url}
\usepackage{verbatim}
\usepackage{graphicx}
\usepackage{cite}
\usepackage{multirow}
\usepackage{graphicx}
\usepackage{caption}
\usepackage{subcaption}
\usepackage{makecell}
\usepackage{booktabs} 

\begin{document}

\title{External Human-Machine Interface based on Intent Recognition: Framework Design and Experimental Validation}

\author{Boya Sun, Haotian Shi, Ying Ni, Shaocheng Jia,~\IEEEmembership{Member,~IEEE,} Haoyang Liang,~\IEEEmembership{Member,~IEEE}
\thanks{The work described in this paper was supported by grants from the National Natural Science Foundation of China (Grant No. 52302378). 
The last author was also supported by Shanghai Super Postdoctoral Incentive Program. 
The experiments in this study were conducted using the TransCAVE platform at Tongji University (http://202.120.189.178:5021/\#/serviceSystem).

Boya Sun, Haotian Shi, Ying Ni and Haoyang Liang are with the College of Transportation and the Key Laboratory of Road and Traffic Engineering, Ministry of Education, Tongji University, Shanghai 201804, China (e-mail: 2410802@tongji.edu.cn, shihaotian95@tongji.edu.cn, ying\_ni@tongji.edu.cn, lianghowie@tongji.edu.cn).

Shaocheng Jia is with the Department of Civil and Environmental Engineering, National University of Singapore, Singapore (e-mail: shaochengjia@nus.edu.sg).

}
}

\markboth{Journal of \LaTeX\ Class Files,~Vol.~X, No.~X, XXXXXXXX}%
{Shell \MakeLowercase{\textit{et al.}}: A Sample Article Using IEEEtran.cls for IEEE Journals}


\maketitle

\begin{abstract}
Increasing autonomous vehicles (AVs) in transportation systems makes effective interactions between AVs and pedestrians indispensable. External human--machine interface (eHMI), which employs visual or auditory cues to explicitly convey vehicle behaviors can compensate for the loss of human-like interactions and enhance AV--pedestrian cooperation. To facilitate faster intent convergence between pedestrian and AVs, this study incorporates an adaptive interaction mechanism into eHMI based on pedestrian intent recognition, namely IR-eHMI. IR-eHMI dynamically detects and infers the behavioral intentions of both pedestrians and AVs through identifying their cooperation states. The proposed interaction framework is implemented and evaluated on a virtual reality (VR) experimental platform to demonstrate its effectiveness through statistical analysis. Experimental results show that IR-eHMI significantly improves crossing efficiency, reduces gaze distraction while maintaining interaction safety compared to traditional fixed-distance eHMI. This adaptive and explicit interaction mode introduces an innovative procedural paradigm for AV--pedestrian cooperation.
\end{abstract}


\section{Introduction}

With the increasing deployment of autonomous vehicles (AVs) on urban roads, interactions between AVs and vulnerable road users (VRUs), including pedestrians, cyclists, and motorcyclists, have become increasingly common, raising significant concerns about VRUs' safety and comfort \cite{b1,b2}. Due to their limited physical protection and high exposure, VRUs are particularly prone to severe injuries in traffic conflicts, with pedestrians representing the most endangered group \cite{b2,b0}. Two major classes of AV--pedestrian interaction issues have been widely reported: dangerous interactions, where pedestrians cannot accurately infer AV intent and consequently face collision risks \cite{itsm-vru,interact}; and inefficient interactions, where both parties hesitate, leading to inefficiencies and ``who-goes-first" deadlocks \cite{interact,communicate}. These issues are mainly due to lack of communication \cite{Shared Spaces,communicate}, which hinders the establishment of trust and smooth cooperation between pedestrians and AVs, thus limiting broader social acceptance.

Generally, there are two primary approaches to address these challenges. The first involves advancing the capabilities of the Autonomous Driving System (ADS) to enhance the safety and comfort of VRUs, by optimizing AV motions through more accurate pedestrian detection or perception methods \cite{ADS1, RN2023RuanGEIT}, as well as more human-like decision-making strategies \cite{RN2025LuAI4T}. However, this approach has limitations. First, ADS cannot communicate with other traffic users, leading to self-centered planning and decisions during interactions. Additionally, the absence of a human driver in AVs eliminates implicit social cues, such as eye contact and hand gestures \cite{communicate,b3}. The second approach compensates these limitations by incoperating external human-machine interface (eHMI), a system that is installed on the exterior of AVs that explicitly communicate the vehicle's status and intentions to nearby road users through visual displays, auditory alerts, or other signaling mechanisms \cite{b3,RN2025ZhangGEIT, eHMI-review}. Its effectiveness on improving AV--pedestrian interactions has been proven through numerous experimental studies \cite{T5,T4,Feng}.

So far, most eHMI merely demonstrates the vehicle's intention to stop or yield \cite{T3,T5,T7}, with AVs largely employing conservative driving strategies. These strategies include driving at lower speeds, avoiding complex interactions, and frequently stopping in uncertain situations to prevent accidents \cite{b7}. While this conservative driving strategy ensures pedestrian safety, it also negatively impacts traffic flow, potentially leading to fuel waste, reduced traffic efficiency, and even ``malicious behavior" (minor bullying AVs) or violations by pedestrians \cite{Shared Spaces}. Weber et al. \cite{b7} warned that, given differences in human cognition and comprehension, in situations where there is a risk of misunderstanding, displaying an eHMI may be unnecessary or even harmful. These findings underscore the importance of deploying eHMI selectively and adaptively, rather than universally. Accordingly, modeling and predicting the intents of interacting agents is essential for developing more intelligent, context-aware eHMI.

To deploy intelligent eHMI systems, a decision-making strategy is needed that can understand pedestrians’ intents based on real-time traffic data and selectively activates eHMI cues. Existing studies for pedestrian behavior modeling generally fall into two categories: rule-based models and data-driven models. Rule-based models offer interpretability but struggle with real-world complexity \cite{SFM, CA, PVG}. In contrast, data-driven models, while effective in prediction, rely on large annotated datasets and often generalize poorly to unexplainable unseen results \cite{PED-RL,ped-irl}. In AV--pedestrian interaction scenarios, pedestrians’ intents are dynamically influenced by the AV's movements, making their behavior context-dependent. Despite advancements in the above models, there is still a lack of efficient and dynamic algorithms specifically designed for recognizing pedestrian intent in AV--pedestrian interactions. As a result, there is a lack of intelligent and explicit communication strategy that can adapt in real-time to the evolving AV--pedestrian interaction.

To bridge this gap, this study proposes IR-eHMI, an adaptive eHMI promoting strategy based on a data-driven IR method. During the dynamic interaction between a pedestrian and AV, IR-eHMI can continuously infer the pedestrian's intent and monitors the dynamic cooperative state to determine whether eHMI prompts should be activated. This methodology addresses the gaps identified above by moving beyond static, one-size-fits-all solutions, offering a more adaptive and context-aware approach to AV--pedestrian interaction. To evaluate its effectiveness, a Virtual Reality (VR)-based experiment is designed and conducted. The evaluation considers both objective behavioral metrics and subjective feedback from participants.

The key contributions of this paper are summarized as follows:
\begin{itemize}
\item[1] This study introduces a context-aware eHMI prompting strategy that selectively promotes eHMI through monitoring the dynamic cooperation state. This approach innovatively adapts eHMI promotion to the changing interaction context between AVs and pedestrians.
\item[2] A VR-based testbed is developed, enabling software-in-the-loop evaluation of context-aware eHMI strategies. This platform offers a novel virtual environment for assessing eHMI's effectiveness.
\item[3] Pedestrian-in-loop (PIL) experimental results provide new insights into how pedestrian heterogeneity impacts eHMI effectiveness: while eHMI improves interaction performance for some users, it may also increase their attentional distraction. 
\item[4] The effectiveness of the proposed IR-eHMI is systematically evaluated through a VR experiment, demonstrating its ability to improve interaction safety and efficiency compared with conventional eHMI designs.
\end{itemize}

\section{Related Work}

\subsection{External Human-Machine Interface}
With the rapid advancement of AV technologies, achieving safe and efficient AV--pedestrian interaction has become a critical challenge. This is particularly pressing for SAE Level $4$ and $5$ AVs, in which human drivers are no longer responsible for vehicle control \cite{SAE}. In such fully autonomous settings, traditional human-to-human communication cues, such as eye contact or hand gestures, are no longer applicable \cite{communicate,b3}. Relying solely on the implicit interaction derived from AV motion (e.g., deceleration or stopping) provides limited communicative capacity, making it difficult for pedestrians to infer vehicle intent. This increases uncertainty and potential safety risks in traffic interactions, ultimately hindering effective AV planning and decision-making.

To address these issues, many studies have proposed the use of eHMI \cite{T4,T8,T6,T9}. EHMI is a communication system mounted on the exterior of AVs, designed to explicitly convey vehicle intentions to pedestrians and other road users through visual, auditory, or multimodal signals \cite{int1}. In recent years, eHMI has evolved to include a wide range of expressive modalities, such as textual messages, icons, dynamic light strips, anthropomorphic symbols, and even ground projections \cite{Feng}. In recent years, an increasing number of empirical studies have evaluated the effectiveness of various eHMI designs and their influence on pedestrian crossing behavior, safety, efficiency, and user acceptance \cite{eHMI-review}. Broadly, these studies can be categorized into two types: survey-based and experiment-based, as summarized in Table \ref{tab_lr}. Survey-based studies have highlighted the potential of eHMI from both pedestrian and AV perspectives, revealing their perceived benefits in enhancing communication and trust. In contrast, experiment-based studies have primarily focused on evaluating different modes and configurations of eHMI, aiming to identify designs that optimize pedestrian understanding and interaction outcomes.

\begin{table*}[htbp]
\caption{Classification of the eHMI-related studies in recent five years.}
\begin{center}
\begin{tabular}{>{\raggedright\arraybackslash}m{4cm}>{\centering\arraybackslash}m{1.5cm}
                >{\centering\arraybackslash}m{2cm}>{\raggedright\arraybackslash}m{4cm}
                >{\raggedright\arraybackslash}m{3.5cm}}
\toprule
\textbf{Reference} & \textbf{Category} & \textbf{Method} &\textbf{Scene} 
& \textbf{eHMI type}\\
 \midrule
 Lanzer and Baumann (2023) \cite{T3}& Survey& Video &Urban road, AV yields&Implicit\\
 
 Harkin, et al. (2023) \cite{T1}& Survey& Video & Intersection, AV yields&Implicit\\
 
 Eisele and Petzoldt (2022) \cite{T2}& Survey& Picture & Zebra crossing, traffic Signals, AV yields / does not yield&Visual icons\\
 
 Lyu, et al. (2024) \cite{T6}& Experiment& Video& Unsignalized intersection, AV yields / does not yield&Distinct eHMI designs with unique combinations\\
 
 Hübner, et al. (2022) \cite{T4}& Experiment& VR &Mixed traffic scenarios, AV yields / does not yield&eHMI for different right-of-way\\
 
 Dey et al. (2021) \cite{T7}& Experiment& VR &Straight section, AV yields&eHMI for concept design, scalable communication\\

 Lau et al. (2024) \cite{T8}& Experiment& VR &Pedestrians crossing, AV yields / does not yield&Dynamic eHMI and static eHMI\\

 Song et al. (2023) \cite{T9}& Experiment& VR &Pedestrians’road-crossing, non-intersection, AV does not yield&eHMI for predicted real-time risk levels\\
 
 Carlowitz, et al. (2023) \cite{T5}& Experiment& Field experiment & Non-intersection, AV yields&Full and partial light band\\

 Alhawiti, et al. (2024) \cite{T10}& Experiment& Field experiment &Intersection, AV yields / does not yield&Red, green and countdown timer\\
 
 Tang et al. (2025) \cite{T11}& Experiment& Field experiment & AV yields / does not yield&Text/symbol/light based eHMI\\ \bottomrule

\end{tabular}

\label{tab_lr}
\end{center}
\end{table*}

Empirical studies have demonstrated that eHMI can enhance pedestrians’ understanding of a vehicle’s yielding intention and reduce decision-making ambiguity—particularly in situations where vehicle speed is low or intent is not clearly perceivable \cite{b6}. For instance, \cite{gaze} found that eHMI significantly influenced pedestrians’ gaze behavior and reaction time. Other studies have shown that eHMI can decrease crossing initiation time (CIT) \cite{lee2023} and enhance pedestrians' willingness to interact with vehicles \cite{will}.

Despite their potential to improve the quality of AV--pedestrian interaction, there remains ongoing debate regarding the necessity \cite{b7} and appropriate timing of eHMI activation. On one hand, several studies suggest that eHMI can increase pedestrians’ perceived safety under specific conditions, such as when the vehicle has stopped or is clearly decelerating. On the other hand, other research has indicated that inconsistencies between vehicle behavior and eHMI signals may result in confusion or distraction, ultimately interfering with pedestrians’ interpretation of motion cues and undermining their sense of safety and trust \cite{b7}. Regarding the timing of eHMI activation, previous research \cite{distance25m} suggests that some pedestrians begin looking at the windshield approximately $25$ meters before the vehicle reaches them while the other also reported a desire for earlier information, highlighting heterogeneity in pedestrians’ expectations of eHMI timing.

Most existing eHMI studies have focused on aspects such as display position, modality, message content, activation distance, and alignment with vehicle kinematics, aiming to improve communication clarity and interaction efficiency. However, the effectiveness of eHMI is not solely determined by how well the message is conveyed, but also by whether it supports pedestrians’ understanding of vehicle intent, especially in ambiguous or complex right-of-way scenarios \cite{T4,T6}. Therefore, designing context-aware eHMI response mechanisms is essential for enhancing the quality and effectiveness of AV--pedestrian interactions.

\subsection{Modeling of AV--Pedestrian Interaction}

Pedestrian behavior and intention are highly unpredictable due to personal, environmental and social factors, posing a significant challenge to AV planning and decision-making \cite{interact}. AV--Pedestrian Interaction modeling seeks to analyze the interaction process, enabling AVs to predict pedestrians' crossing intention and make appropriate decisions. Models can be broadly categorized into two categories: rule-based models, such as Social Force Models (SFM) \cite{SFM}, Cellular Automata (CA) models \cite{CA} and Pedestrian--Vehicle Game (PVG) models \cite{PVG}; and data-driven models, such as classifier-based methods \cite{SVM,RN2025LiuGEIT} and Reinforcement Learning (RL)-based methods \cite{PED-RL, ped-irl, 2024YeRLV2P}.

Rule-based models rely on explicitly defined rules to govern pedestrians’ state transitions and decision-making processes \cite{track}. Anvari et al. \cite{SFM} developed a three-layer microscopic model to simulate the interaction between pedestrians and vehicles in shared spaces, and modified and extended the Social Force Model to generate feasible movements. Zhang et al. \cite{CA} developed a simulation model that integrated the strengths of cellular automata with probabilistic functions, providing a realistic mechanism to capture the competitive and conflicting interactions between vehicle and pedestrian flows. And Wu et al.  \cite{PVG} proposed a game theory–based interaction model, calibrated with empirical video data and integrated into a cellular automaton traffic simulation, to analyze the influence of pedestrian crossing behavior on traffic performance at unsignalized crosswalks. Despite their interpretability, such models often struggle to capture the full complexity of real-world pedestrian behavior and suffer from limited scalability due to their reliance on assumed rules.

Data-driven models learn interaction and behavior patterns from real data, and are thus able to capture more complex features. Ni et al. \cite{SVM} used the Support Vector Machine (SVM) approach to classify the severity of interaction events extracted from intersections and analyzed the importance of variables. Wu et al. \cite{itsm-intent-prediction} developed the Fast Road User Interaction (FRUIT) framework to efficiently model road user interactions by combining latent intention estimation and multimodal trajectory prediction, improving AV performance in complex traffic scenarios. Wang et al. \cite{PED-RL} introduced a reinforcement learning model with constraints of human sensory and motor to simulate the crossing decision and locomotion of pedestrians. Nasernejad et al. \cite{ped-irl} proposed a Gaussian Process Inverse Reinforcement Learning (GP-IRL) approach to retrieve pedestrians’ reward functions and infer their collision avoidance mechanisms in conflict situations. 

These models typically focus on predicting or replicating strategies and actions, but often lack mechanistic explanations for the interaction processes. Similar to AV-HV interactions \cite{IR-consis}, pedestrian intent evolves dynamically throughout the interaction, continuously influenced by the AV’s behavior and the surrounding context, and ultimately converges toward a shared consensus. Therefore, it is essential to efficiently and dynamicly model pedestrian intent and its coupling with AV's intent to enable more interpretable and context-aware interaction communication.

\section{Methodology}

\begin{figure}[htbp]
  \centering
  \includegraphics[width=0.99\linewidth]{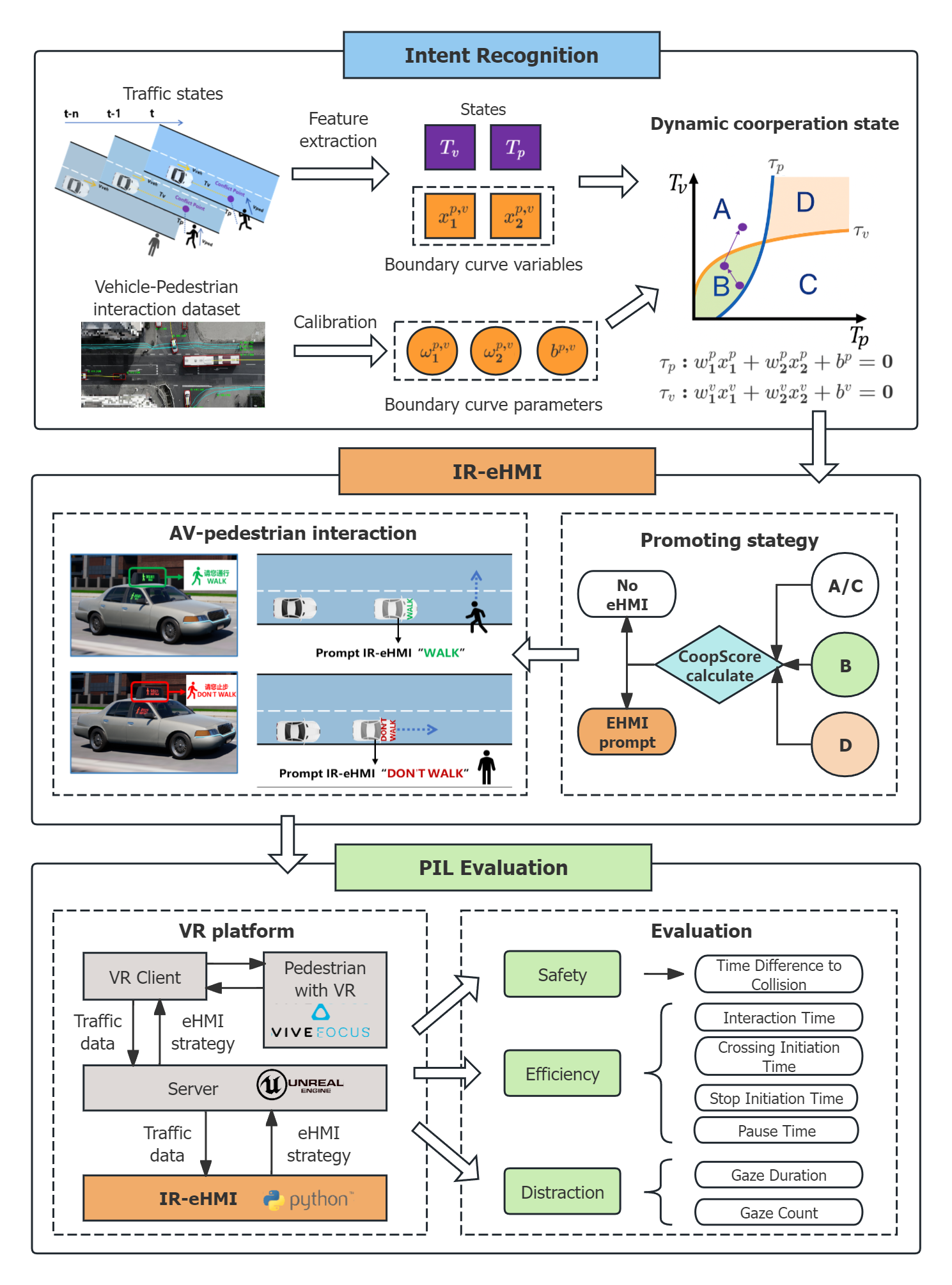}
  \caption{Overall framework of IR-eHMI and evaluation}
  \label{fig_framework}
\end{figure}

IR-eHMI incorporates an Intent-Recognition-based external human-machine interface (IR-eHMI) to facilitate adaptive and explicit interactions between AVs and pedestrians. It aims to improve AV--pedestrian communication by monitoring the dynamic cooperation states and selectively promoting eHMIs. Fig. \ref{fig_framework} shows the overall framework for mechanism design and validation, and the inputs and functions of the modules are as follows:

\textbf{Intent Recognition}. This stage achieves IR through analyzing naturalistic vehicle--pedestrian interaction data and calibrating boundary curve parameters on the dynamic cooperation state. The input is interaction data, and the output is the calibrated parameters for monitoring the dynamic cooperation state.

\textbf{IR-eHMI}. Based on the calibrated parameters and real-time data, this stage monitors the cooperation state and determines whether to promote eHMIs to improve interaction. The input is continuous AV--pedestrian interaction data, and the output is the updated cooperation state and eHMI promotion strategy.

\textbf{PIL Evaluation}. The final stage evaluates the performance of the IR-eHMI mechanism on a VR-based platform. The input is the deployable IR-eHMI software, and the output is performance indicators related to safety, efficiency, and distraction.

\subsection{Intent recognition}

Pedestrian intent is not a static or instantaneous event, but a dynamic process influenced by the motion states of both parties and their tolerance for different behavioral outcomes, eventually converging towards consensus. To efficiently recognize the intent of pedestrians and identify characteristics on the dynamic cooperation state, the concept of cooperative acceleration \cite{IR-consis} is introduced, which is defined as the minimum acceleration required for a pedestrian to just avoid collision with the interacting vehicle, calculated as:

\begin{equation}
a_{c,\text{int}} = \frac{2\left(d_{\text{int}} - v_{\text{int}}T_{\text{self}}\right)}{T_{\text{self}}^2}
\end{equation}

where:
\begin{itemize}
  \item Subscript \textit{self} denotes the ego agent (AV or pedestrian),
  \item Subscript \textit{int} denotes the interacting agent,
  \item $T_{\text{self}}$ is the ego agent's time to reach the conflict point,
  \item $d_{\text{int}}$ is the distance from the interacting agent's current position to the conflict point,
  \item $v_{\text{int}}$ is the interacting agent's current velocity.
\end{itemize}

The key idea is that agents evaluate how their actions will influence the other party, particularly whether this influence exceeds the other's acceptable behavioral threshold—thus shaping mutual anticipation of intent. Based on this, cooperative acceleration offers two interpretive cases:

\begin{itemize}
  \item When $a_{c,\text{ped}} > 0$ and is large, the pedestrian must accelerate aggressively to reach the conflict point simultaneously with the vehicle, thus tending to yield and cross after the AV.
  \item When $a_{c,\text{ped}} < 0$ and is large in magnitude, the pedestrian must decelerate significantly or stop, tending to cross before the AV arrives.
\end{itemize}

While cooperative acceleration provides insights into momentary intention and its strength, its precise relationship with final decision outcomes remains uncertain. Noting its structural properties: (1) The numerator reflects the direction of intention. (2) The denominator reflects the magnitude and sensitivity of the tendency. Thus, we define:

\begin{subequations}
\begin{align}
& x_1 = T_{\text{self}}^2\\
&x_2= 2(d_{\text{int}} - v_{\text{int}}T_{\text{self}}) 
\end{align}
\end{subequations}

\begin{table*}[t!]
\caption{SVM Classification Results}
\centering
\label{tab:svm_results}
\begin{tabular}{ccccccccc}
\toprule
\textbf{Self} & \textbf{Interact} & $\boldsymbol{\omega_1}$ & $\boldsymbol{\omega_2}$ & $\boldsymbol{b}$ 
& \textbf{Accuracy} & \textbf{Precision} & \textbf{Recall} & \textbf{F1 score} \\
\midrule
Ped & AV & -0.0032 & 0.0469 & 0.2503 & 95.12\% & 93.94\% & 93.94\% & 93.94\% \\
AV & Ped & -0.0288 & 0.1769 & 0.7601 & 96.34\% & 95.39\% & 98.64\% & 96.99\% \\
\bottomrule
\end{tabular}
\end{table*}

Each pedestrian--vehicle interaction trajectory is discretized into multiple sample points. For each point, $(x_1, x_2)$ are used as input features, and the ground truth behavior (i.e., whether the pedestrian crosses first or yields) serves as the label. Due to the relatively small dataset size and the need for model robustness, a Support Vector Machine (SVM) is employed to establish the classification boundary:

\begin{equation}
f_{\text{SVM}}: \omega_1 x_1 + \omega_2 x_2 + b = 0
\end{equation}

where $\mathbf{W} = [\omega_1, \omega_2, b]$ are the learned boundary curve parameters. 

Obviously:
\begin{itemize}
 \item Since $\omega_1 < 0$, the boundary curve is an upward-opening parabola.
 \item Given $b > 0$ and $\Delta = 1 - \frac{\omega_1 b}{(\omega_2 v_{\text{int}})^2} > 0$, a single intersection exists in the positive axis region.
\end{itemize}

Before applying the proposed IR method to a specific scenario, the boundary curve parameters $\mathbf{W}$ need to be calibrated. To achieve this, the study utilizes the inD dataset, which provides naturalistic traffic behavior data captured by drones at unsignalized urban intersections \cite{b9}. 

To extract relevant pedestrian--vehicle interactions, trajectory segments involving strong interactions were identified based on the following criteria:
\begin{itemize}
 \item The absolute value of the time difference to collision ($\vert$TDTC$\vert$) between vehicle and pedestrian is less than $3$ $\mathrm{s}$, and
 \item The minimum spatial distance between them is less than $5$ $\mathrm{m}$.
\end{itemize}
\begin{equation}
|\text{TDTC}| = \left| T_{\text{self}} - \frac{d_{\text{int}}}{v_{\text{int}}} \right|
\end{equation}
Applying this, a total of $275$ pedestrian--vehicle interaction segments were extracted for SVM training. The calibrated boundary curve parameters derived from different agents are summarized in Table~\ref{tab:svm_results}.

\subsection{IR-eHMI}
Based on the inputs of the real-time traffic states, cooperation state between the pedestrian and the vehicle could be dynamically updated. In the previous work, we explored the expected convergence direction based on the current state from a unilateral perspective, but intent consistency is the result of the convergence of bilateral intentions in integrated interactions. Therefore, this module uses graphical methods to couple the expected convergence directions of both parties' intentions. The two boundary curves derived from the previous module divide the feature space into four distinct regions \cite{IR-consis}, representing different cooperation states, as illustrated in Fig.~\ref{fig_framework}:

\begin{itemize}
 \item \textbf{Region A (Intent Convergence)}: Pedestrian prefers to go first, while AV yields. Mutual agreement achieved.
 \item \textbf{Region B (Intent Conflict I)}: Both parties prefer to yield. Risk of mutual hesitation.
 \item \textbf{Region C (Intent Convergence)}: Pedestrian yields, while AV proceeds first. Mutual agreement achieved.
 \item \textbf{Region D (Intent Conflict II)}: Both parties prefer to proceed first. Competitive interaction possible but manageable.
\end{itemize}

When the interaction state approaches the boundary of decision regions, the behavioral intentions of pedestrians and vehicles often exhibit ambiguity and uncertainty. In such cases, binary classification may lead to misjudgments. Furthermore, due to the inherent delay and continuity in human and vehicle behavior during real-world interactions, the hard judgment fails to capture the gradual variation in ``intent consistency", limiting the granularity of interaction representation. 

To address this issue, we introduce a continuous metric termed the Cooperation Score, aiming to quantify the degree of intent consistency between the pedestrian and the AV at each time frame. The Cooperation Score is computed based on the current cooperation states $(T_p, T_v)$ and the two decision boundary curves $\tau_v(T_p)$ and $\tau_p(T_v)$. We define the discriminant distances as the deviations from the respective boundaries:

\begin{subequations}
\begin{align}
&d_v = T_v - \tau_v(T_p),\\
&d_p = T_p - \tau_p(T_v).
\end{align}
\end{subequations}

$(d_p, d_v)$ represent the deviations of the vehicle's and pedestrian's conflict arrival times from the critical thresholds, respectively. To map these distances to a normalized scale, we apply the sigmoid function $\sigma(\cdot)$: $s_v = \sigma(d_v),s_p = \sigma(d_p)$, and the Cooperation Score is then defined as:

\begin{equation}
\text{CoopScore} = s_v \cdot (1 - s_p) + s_p \cdot (1 - s_v)
\end{equation}

The Cooperation Score captures the essential characteristic of intent conflicts and convergence: When $s_v$ and $s_p$ are close in value (e.g., both approaching $0$ or $1$), it indicates mutual hesitation or competition, leading to a lower CoopScore. In contrast, when the values are clearly opposite (e.g., one high and one low), it reflects strong complementary behavior, resulting in a higher CoopScore.

In real-time implementation, the CoopScore is updated continuously on a frame-by-frame basis and used to determine whether eHMI prompts should be displayed:

\begin{itemize}
 \item When the interaction lies in Region A/C ($d_v \cdot d_p < 0$) and $\text{CoopScore} \geq 0.9$, it indicates high intention alignment, and no eHMI is triggered to avoid unnecessary distractions.
 \item When the interaction lies in Region B ($d_v < 0$, $d_p < 0$) and $\text{CoopScore} < 0.9$, it suggests mutual hesitation, and an eHMI prompt is needed to improve efficiency.
 \item When the interaction lies in Region D ($d_v > 0$, $d_p > 0$) and $\text{CoopScore} < 0.9$, it implies mutual intention to proceed first, and an eHMI prompt is needed to enhance safety.
\end{itemize}

\subsection{PIL evaluation}

\begin{figure*}[b!]
 \centering
 \includegraphics[width=\linewidth]{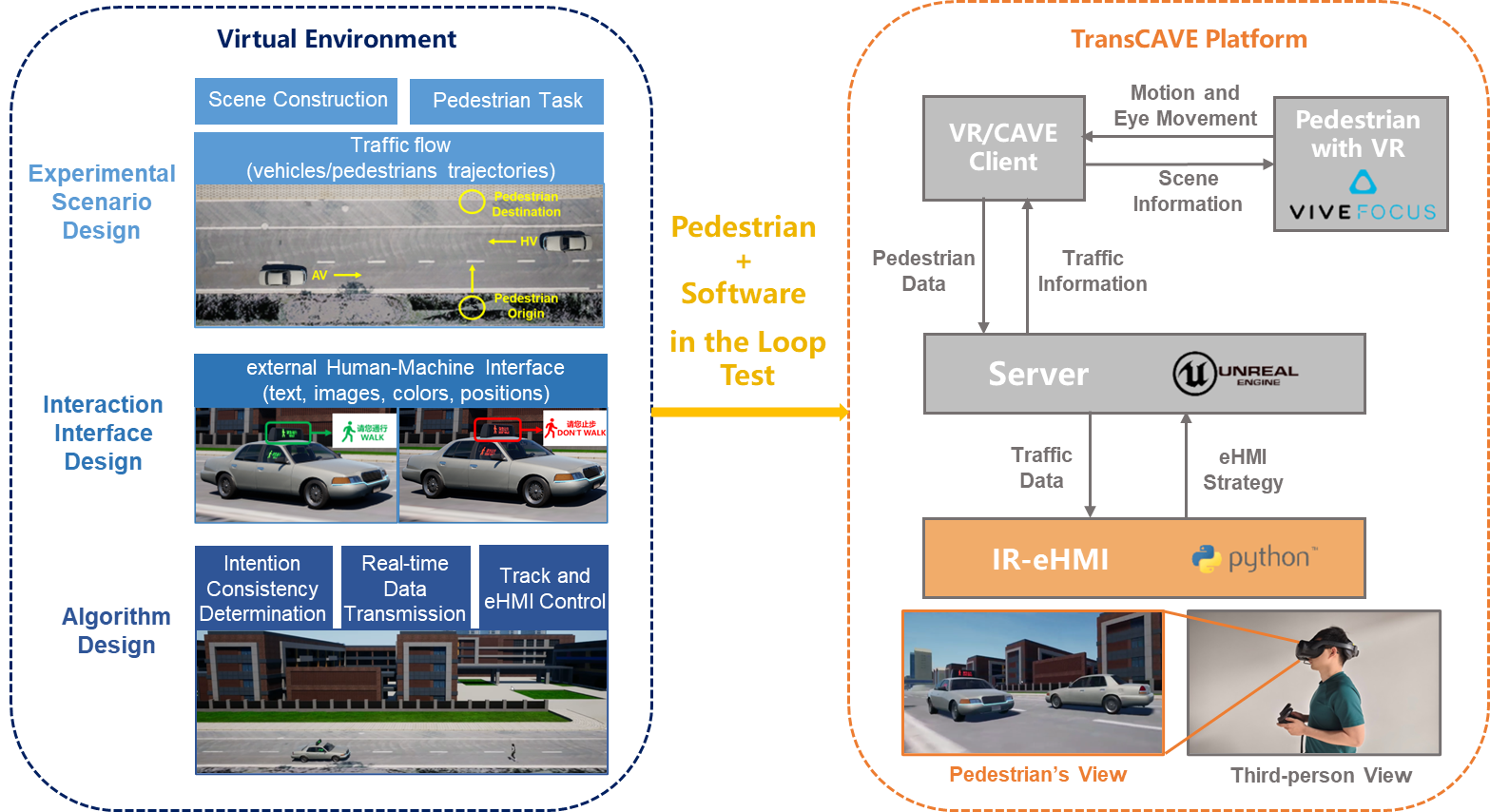}
 \caption{Implementation of IR-eHMI on TransCAVE. }
 \label{PIL}
\end{figure*}

The proposed IR-eHMI is then applied to a VR-based experimental platform supported by TransCAVE Laboratory at Tongji University. One of its functions is to design, evaluate, and optimize eHMI for AVs through digital-twin-based human–vehicle interaction scenarios, as shown in Fig.~\ref{PIL}. The platform deeply integrates vehicles, pedestrians, and traffic simulation systems to construct a reproducible, controllable, and evaluable virtual traffic environment, enabling us to analyze various interaction scenarios in a safe, flexible, and realistic setting. On TransCAVE, we design experimental scenarios and eHMI interfaces, and implement the IR-eHMI algorithm as illustrated in Fig. \ref{PIL}. 

Through real-time data transmission between the Unreal Engine server and VR clients, the platform enables both pedestrian-in-loop and software-in-loop testing environment, in which the pedestrian and the virtual AV equiped with IR-eHMI interact dynamically and synchronously. The key components are described as follows:

\begin{itemize}
 \item The \textbf{Server} builds and runs the environment, background traffic, and interaction scenarios, using Unreal Engine.
 \item The \textbf{Client} is connected to the VR equipment, which presents visual and auditory information of virtual vehicles and the surrounding environment to the pedestrian.
 \item The \textbf{Pedestrian} interacts with the virtual environment through a VR headset (HTC VIVE Focus 3), using a handheld controller to navigate walking direction and speeds.
 \item When the intent conflict is identified by the \textbf{IR-eHMI} software module, an appropriate eHMI prompt (e.g., ``WALK" or ``DON'T WALK") is displayed on the AV interface, as illustrated in Fig.~\ref{fig_framework}.
\end{itemize}

The platform supports the collection of multiple data modalities, including video recordings, trajectories, and eye-tracking data. After conducting data cleaning and preprocessing, the evaluation is carried out across several aspects, with objective indicators including safety (Time difference to collision \cite{TDTC}), crossing efficiency (Interaction Time, Crossing Initiation Time, Stop Initiation Time, Hesitation Time) \cite{eHMI-review}, and distraction (Gaze duration, Gaze Count) \cite{gaze}. The definitions of the indicators are as follows:
\begin{itemize}
 \item \textbf{$\vert$TDTC$\vert$ (Absolute value of Time Difference To Collision)}: the absolute value time of the difference for pedestrian and vehicle travelling to the potential conflict point if their speed and direction keep constant.
 \item \textbf{IT (Interaction Time)}: The time from when the pedestrian first gazes at the vehicle until either the pedestrian or the vehicle fully exits the potential conflict zone.
 \item \textbf{CIT (Crossing Initiation Time)}: The time from when the pedestrian first gazes at the vehicle until they start crossing the road continuously, serving as a measure of intent consistency in AV-yielding scenarios.
 \item \textbf{SIT (Stop Initiation Time)}: The time from when the pedestrian first gazes at the vehicle until they begin to stop and yield, serving as a measure of intent consistency in AV non-yielding scenarios.
 \item \textbf{HT (Hesitation Time)}: The duration from interaction onset to the start of continuous crossing, during which the pedestrian pauses (speed $<$ $0.5$ $\mathrm{m/s}$ \& deceleration).
 \item \textbf{GD (Gaze Duration)}: The total time the pedestrian gazes at the vehicle.
 \item \textbf{GC (Gaze Count)}: The number of distinct times the pedestrian gazes at the vehicle.
\end{itemize}

\section{Experimental Evaluation}

\subsection{Experimental design}

\begin{figure}[htbp]
 \centering

 \begin{subfigure}[b]{0.99\linewidth}
  \centering
  \includegraphics[width=\linewidth]{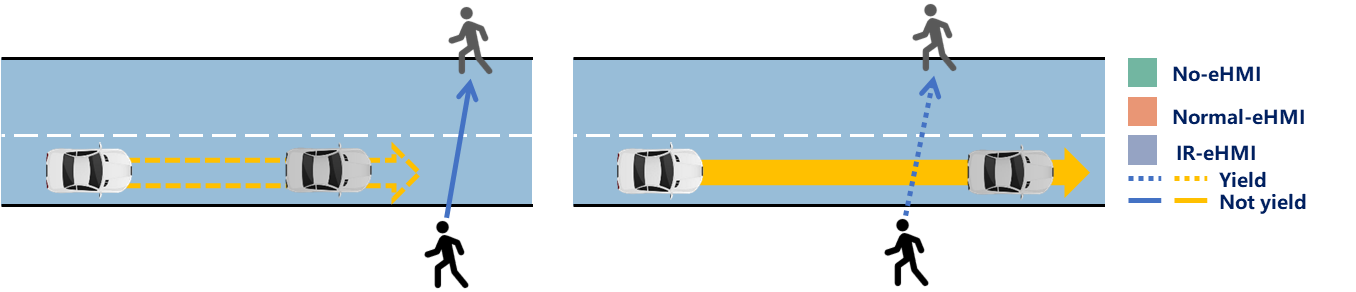}
  \caption{Overall experiment design}
  \label{fig_overall_exp}
 \end{subfigure}
 \vskip\baselineskip

 \begin{subfigure}[b]{0.49\linewidth}
  \centering
  \includegraphics[width=\linewidth]{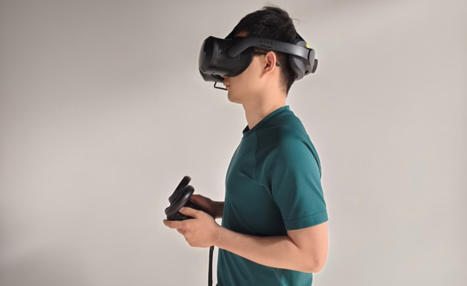}
  \caption{Pedestrian with VR}
  \label{fig_ped_VR}
 \end{subfigure}
 \hfill
 \begin{subfigure}[b]{0.49\linewidth}
  \centering
  \includegraphics[width=\linewidth]{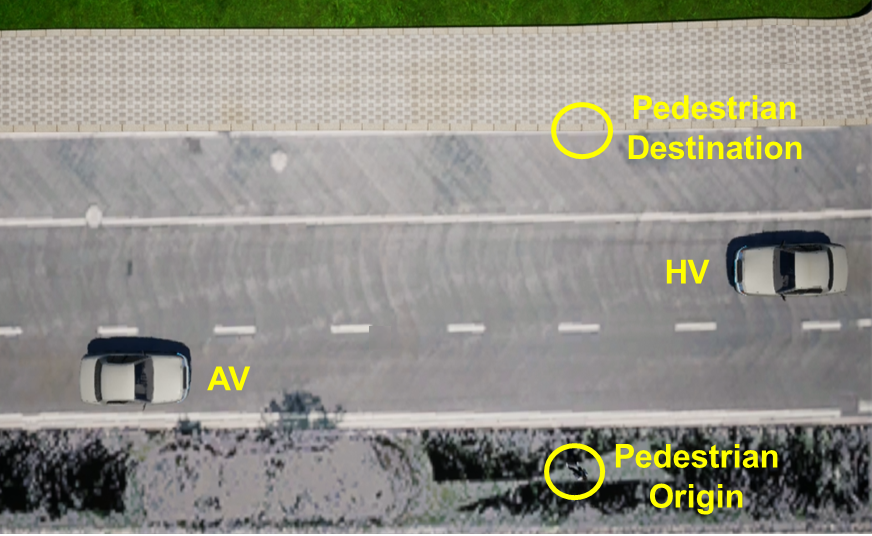}
  \caption{Bird view}
  \label{fig_bird_view}
 \end{subfigure}
 \vskip\baselineskip

 \begin{subfigure}[b]{0.49\linewidth}
  \centering
  \includegraphics[width=\linewidth]{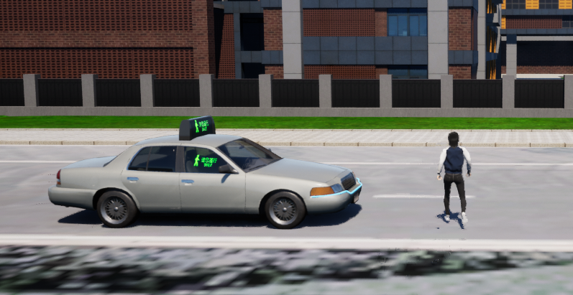}
  \caption{S1: AV yielding}
  \label{fig_S1}
 \end{subfigure}
 \hfill
 \begin{subfigure}[b]{0.49\linewidth}
  \centering
  \includegraphics[width=\linewidth]{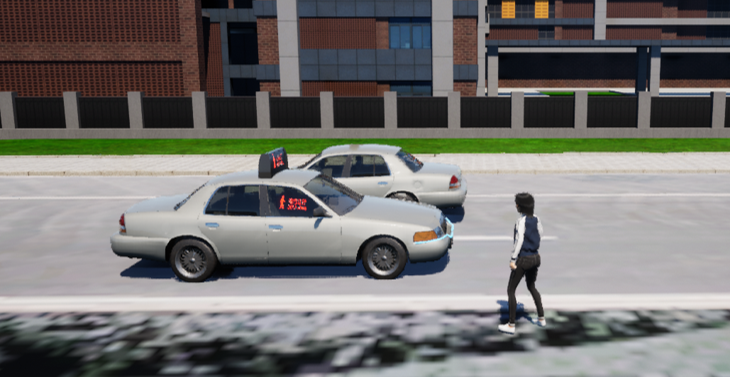}
  \caption{S2: AV non-yielding}
  \label{fig_S2}
 \end{subfigure}

 \caption{Experiment design of the PIL evaluation.}
 \label{fig_scenario_overview}
\end{figure}

To assess the practical effectiveness of the proposed IR-eHMI and evaluate its generalizability, a VR experiment was conducted under two distinct interaction scenarios, denoted as S1 and S2, using the TransCAVE platform. This human-involved experiment was conducted with informed consent and received approval from the Ethics Committee at Tongji University (No. tjdxsr2024041). 
Controlled experiments were conducted including No-eHMI, Normal-eHMI (prompt based on fixed-distance threshold \cite{b6}) and the proposed IR-eHMI. The scenario setups and corresponding eHMI cues for each condition are illustrated in Fig.~\ref{fig_scenario_overview}.

\begin{table*}[ht]
\centering
\footnotesize
\caption{Summary of Experimental Scenarios and eHMI Modes}
\label{tab:experiment_summary}
\begin{tabular}{
>{\centering\arraybackslash}p{1.2cm}
>{\centering\arraybackslash}p{2.4cm}
>{\centering\arraybackslash}p{2.4cm}
>{\centering\arraybackslash}p{3.2cm}
>{\centering\arraybackslash}p{2.5cm}
}
\toprule
\textbf{ID} & \textbf{Scenario} & \textbf{eHMI} & \textbf{Prompt condition} & \textbf{eHMI Message} \\
\midrule
S1-A & AV yielding   & No-eHMI   & -         & - \\
S1-B & AV yielding   & Normal-eHMI & Distance $\leq$ 25 m & WALK \\
S1-C & AV yielding   & IR-eHMI   & Intent conflict  & WALK \\
S2-A & AV non-yielding & No-eHMI   & -         & - \\
S2-B & AV non-yielding & Normal-eHMI & Distance $\leq$ 25 m & DON'T WALK \\
S2-C & AV non-yielding & IR-eHMI   & Intent conflict  & DON'T WALK \\
\bottomrule
\end{tabular}
\end{table*}

\subsubsection{S1: AV yielding scenario }

  On a two-way, two-lane road, the AV starts $32$ meters away from the pedestrian, traveling at $7$ m/s in the outer lane. It begins to decelerate at $15$ meters and comes to a complete stop at $2.5$ meters before reaching the pedestrian. A human-driven vehicle (HV) is approaches from the opposite lane at a speed of $7$ m/s. Three controlled experiments were conducted under this setting, where the AV's movement profile remained the same across experiments, and the eHMI mode varied:
  \begin{itemize}
    \item No-eHMI.
    \item Normal-eHMI: Fixed-distance eHMI prompt that display ``WALK'' when the AV is within $25$ meters of the pedestrian.
    \item IR-eHMI: Intent-recognition-based eHMI prompt that display ``WALK'' when an intent conflict is detected and $\text{CoopScore} < 0.9$.
  \end{itemize}

\subsubsection{S2: AV non-yielding scenario }

  On a two-way, two-lane road, the AV starts $35$ meters away from the pedestrian and drives through at a constant speed of $7$ m/s without yielding. A human-driven vehicle (HV) approaches from the opposite lane at a speed of $7$ m/s. Three controlled experiments were conducted under this setting, where the AV's motion remained consistent, and the eHMI mode varied:
  \begin{itemize}
    \item No-eHMI.
    \item Normal-eHMI: Fixed-distance eHMI prompt that display ``DON'T WALK'' when the AV is within $25$ meters of the pedestrian.
    \item IR-eHMI: Intent-recognition-based eHMI prompt that display ``DON'T WALK'' when an intent conflict is detected and $\text{CoopScore} < 0.9$.
  \end{itemize}

\subsection{Procedure}

The experiments recruited $32$ participants, and each participant signed an informed consent form and were briefed on the experimental tasks and requirements. They were informed that the vehicle closer to the pedestrian lane is an AV, while the vehicle in the outer lane is a HV. The behavior of AV in each scenario would vary (it might yield or not yield) and that the AV might or might not display eHMI information via messages on its rooftop or windows. Participants need to cross the road as quickly as possible, while avoiding collisions with vehicles. Prior to the formal experiment, each participant wore VR glasses (Fig.~\ref{fig_ped_VR}) and underwent eye-tracking calibration, and completed three practice crossings to familiarize themselves with the virtual environment. All participants signed the consent form agreeing to take part in the study, and completed the first two parts of the questionnaire. The first section gathered demographic information, including gender, age, and educational background. The second section collected experiential information, including the level of understanding of AVs, road-crossing habits, and prior exposure to eHMI. 

The formal experiment consisted of six test conditions, as summarized in Table~\ref{tab:experiment_summary}. The order of these experimental modules was randomized for each participant to mitigate familiarity effects. Participants were required to complete the corresponding questionnaire to evaluate their cognition and perception during the interaction process \cite{questionnaire}, mainly focusing on the following aspects. Participants received corresponding compensation after the experiments.

perceived safety, recognition of the surrounding environment, comprehensibility of the eHMI, and factors influencing their decision to cross the street. Participants received corresponding compensation after the experiment.

\begin{itemize}
 \item  \textbf{Q-Safe}: Questionnaire score of perceived security, ranging from 0 to 10 (from low to high).
 \item \textbf{Q-Int}: Questionnaire score for the timing of recognizing AV’s intention, ranging from 0 to 10 (from late to early). A higher score indicates that the pedestrian is able to recognize AV’s intent (e.g., to yield or not yield) at an earlier stage during the interaction.
 \item \textbf{Q-Env}: Questionnaire score for the attention to the traffic environment, ranging from 0 to 10  (from low to high). This indicates whether the pedestrian is paying attention to their surroundings.
 \item \textbf{Q-eHMI}: Questionnaire score for perceived eHMI effectiveness if activated, ranging from 0 to 10 (from low to high).

\end{itemize}

\subsection{Results}


\subsubsection{In S1, IR-eHMI improves efficiency}

\begin{table*}[!htbp]
\centering
\caption{Result from one-way ANOVA in S1 (*$p < 0.05$, **$p < 0.01$, ***$p < 0.001$)}

\begin{tabular}{ccccccccc}
\toprule
\textbf{eHMI Type} &
\textbf{IT($\mathrm{s}$)}&
\textbf{CIT($\mathrm{s}$)}&
\textbf{GD($\mathrm{s}$)}&
\textbf{GC} &
\textbf{HT($\mathrm{s}$)}&
\textbf{Q-Int}&
\textbf{Q-Safe}&
\textbf{Q-Env}\\
\midrule

No-eHMI &
10.06$\pm$2.45 &
6.90$\pm$2.45 &
5.79$\pm$2.26 &
2.41$\pm$0.76 &
5.62$\pm$2.47 &
5.09$\pm$2.29 &
6.44$\pm$1.44 &
7.31$\pm$2.04 \\

Normal-eHMI &
8.77$\pm$2.40 &
5.44$\pm$2.52 &
5.04$\pm$1.67 &
2.13$\pm$0.71 &
4.23$\pm$2.36 &
6.63$\pm$1.72 &
7.47$\pm$1.19 &
6.25$\pm$1.34 \\

IR-eHMI &
7.63$\pm$2.01 &
4.79$\pm$1.79 &
4.18$\pm$1.94 &
1.78$\pm$0.66 &
3.45$\pm$2.15 &
7.41$\pm$1.70 &
7.66$\pm$1.21 &
7.00$\pm$1.65 \\

\midrule
$p$&
0.000$^{***}$ &
0.001$^{**}$ &
0.006$^{**}$ &
0.003$^{**}$ &
0.001$^{**}$ &
0.000$^{***}$ &
0.000$^{***}$ &
0.041$^{*}$ \\
\bottomrule

\end{tabular}
\label{tab:ANOVA_S1}
\end{table*}

\begin{figure}[htbp]
  \centering
  \includegraphics[width=\linewidth]{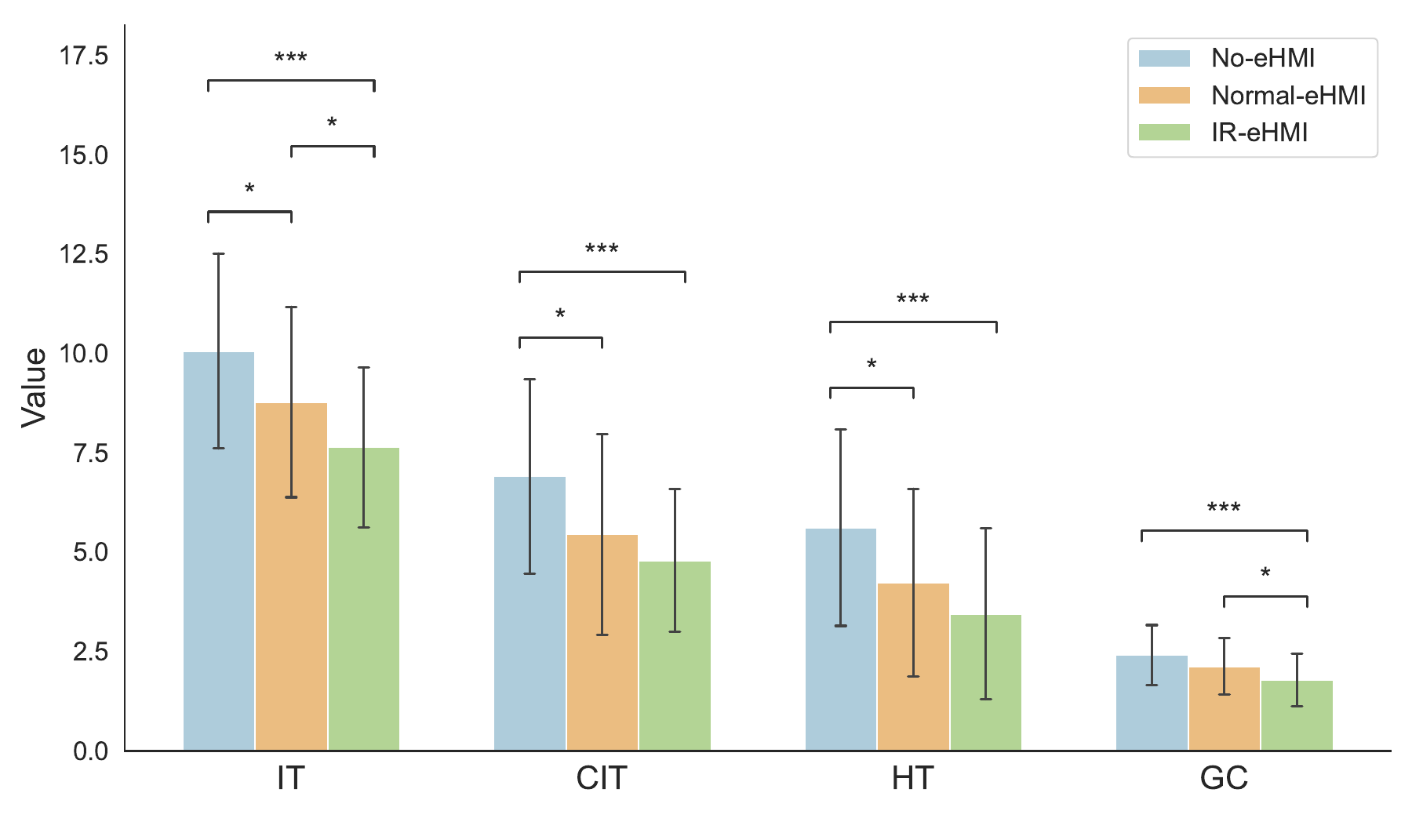}
  \caption{Comparison of IT, CIT, HT, and GC in S1 (*$p < 0.05$, **$p < 0.01$, ***$p < 0.001$).}
  \label{fig_result_case1}
  
\end{figure}

In \textbf{S1} scenario, we used one-way ANOVA to statistically analyze subjective and objective indicators. The effectiveness of different eHMI types can be reflected through pedestrian interaction behavior. One-way ANOVA was used to identify group-level differences (Table\ref{tab:ANOVA_S1}), followed by pairwise independent t-tests to determine specific significant differences. The results are visualized and compared in Fig. \ref{fig_result_case1}. 

Under the \textbf{No-eHMI} condition, pedestrians judged intent solely by vehicle motions. They exhibited the longest average IT $(10.06 \pm 2.45$ $\mathrm{s})$, CIT $(6.90 \pm2.45$ $\mathrm{s})$, GD $(5.79 \pm 2.26$ $\mathrm{s})$ and HT $(5.62 \pm 2.47$ $\mathrm{s})$ compared with the other two eHMI types. In terms of subjective evaluation, Q-Int $(5.09 \pm 2.29$ $\mathrm{s})$ and Q-Safe $(6.64 \pm 1.44$ $\mathrm{s})$ were lower, but Q-Env $(7.31 \pm 2.04$ $\mathrm{s})$ was higher. These results demonstrate that under No-eHMI, pedestrians recognize the AV's yielding intention later, experience a significantly reduced sense of security, and show more attention to the traffic environment. This indicates that, in the absence of external cues, pedestrians require more time to infer the vehicle’s intention and tend to adopt a more cautious crossing strategy. 

Under the \textbf{Normal-eHMI} condition, pedestrian efficiency generally improved. Average IT decreased by approximately $12.8\%$ compared to the \textbf{No-eHMI} condition, while HT and CIT also showed reductions $(p<0.05)$. Moreover, Normal-eHMI introduced greater inter-individual variability across IT $(8.77 \pm 2.40$ $\mathrm{s})$ and CIT $(5.44 \pm 2.52$ $\mathrm{s})$ , suggesting that while some pedestrians benefited from the added information, others may experience hesitation. Meanwhile, the lower Q-Env $(6.25 \pm 1.34$ $\mathrm{s})$ indicated a decrease in pedestrian attention to the surrounding traffic environment. These findings imply that fixed-distance-based eHMI may create inconsistent outcomes and, in some cases, hinder rather than support the interaction.

The \textbf{IR-eHMI} condition demonstrated the most significant benefits. By providing prompts based on real-time interaction state and intent consistency during critical phases of the crossing process, IR-eHMI enabled pedestrians to clearly understand the AV's intent earlier $(7.41\pm 1.40$) and decide crossing faster, achieving the highest efficiency from IT $(7.63 \pm 2.01$ $\mathrm{s})$, CIT $(4.79 \pm 1.79$ $\mathrm{s})$, HT $(2.63 \pm 1.99$ $\mathrm{s})$ and the least visual gaze from GD $(4.18\pm 1.94$ $\mathrm{s})$ and GC $(1.78\pm 0.66$). Furthermore, IR-eHMI resulted in higher pedestrians' perceived safety $( \text{Q-Safe = }7.66\pm 1.21)$ while maintaining a high level of attention to the surrounding traffic environment $(\text{Q-Env = }7.00\pm 1.65$). This demonstrated the potential of dynamic, interaction-driven eHMI signals in conveying AV intent, improving crossing efficiency and reducing attentional distraction.

In summary, the results indicated that pedestrians rely heavily on AV motion cues and exhibited high uncertainty about AV's intentions. Although Normal-eHMI could improve interaction efficiency by reducing IT by $12.8\%$, it did not significantly reduce inter-individual variability. In contrast, IR-eHMI enabled pedestrians to more efficiently interpret the AV’s intent and achieved a $2.5\%$ increase in perceived security, thus performing best in both efficiency and safety dimensions.

\subsubsection{In S2, IR-eHMI maintains safety}

\begin{table*}[b!]
\centering
\caption{Result from one-way ANOVA in S2 (*$p < 0.05$, **$p < 0.01$, ***$p < 0.001$).}
\label{ANOVA_S2}
\renewcommand{\arraystretch}{1.2}
\begin{tabular}{ccccccc}
\toprule
\textbf{eHMI Type} & 
\textbf{\makecell{SIT ($\mathrm{s}$)}}& 
\textbf{\makecell{GC}}& 
\textbf{Q-Int}& 
\textbf{Q-Safe}& 
\textbf{Q-Env}& 
\textbf{Q-eHMI}\\
\midrule
No-eHMI    & 3.04$\pm$1.64 & 1.55$\pm$0.67 & 5.77$\pm$1.90 & 5.75$\pm$1.62 & 7.11$\pm$1.99 & -- \\
Normal-eHMI  & 2.96$\pm$1.49 & 2.00$\pm$0.66 & 7.35$\pm$1.50 & 6.84$\pm$1.68 & 5.97$\pm$1.99 & 7.42$\pm$1.78 \\
IR-eHMI    & 2.17$\pm$1.12 & 1.67$\pm$0.62 & 7.67$\pm$1.24 & 7.33$\pm$0.91 & 7.17$\pm$1.50 & 8.44$\pm$1.38 \\
\midrule
$p$& 0.036* & 0.038* & 0.000*** & 0.000*** & 0.025* & 0.041* \\
\bottomrule
\end{tabular}
\end{table*}

In \textbf{S2} scenario, interactive safety and pedestrian's perceived safety were the primary evaluation focus. Results revealed that under the \textbf{IR-eHMI} condition,the eHMI was activated in $17$ out of $32$ trials, compared with $100\%$ activation in the S1 scenario. This indicated under the AV non-yielding condition, pedestrians were generally able to infer vehicle intent more clearly without explicit interaction. 

Table \ref{ANOVA_S2} shows the subjective and objective indicators and their ANOVA significance test results for different eHMI types. Significant differences were observed across all key indicators, including SIT, GC, Q-Int, Q-Env, Q-Safe and Q-eHMI $(p<0.05)$.

Under the \textbf{No-eHMI} condition, pedestrians exhibited the longest average SIT $(3.04\pm 1.64$ $\mathrm{s})$ with relatively wide distributions. Although the average GC $(1.55 \pm 0.67)$ was the least among all conditions, it showed substantial variability, reflecting differing levels of visual attention and interpretation strategies for implicit interaction. Due to the lack of external cues, pedestrians paid more attention to the surrounding traffic environment $(\text{Q-Env = }7.11 \pm 1.99)$, and the timing of their clear recognition of the AV's non-yielding intention was later $(\text{Q-Int = }5.77 \pm 1.90)$, which increased the uncertainty and perceived risk in the interaction.

Under the \textbf{Normal-eHMI} condition, pedestrians' understanding of the AV's intentions improved. Although the SIT $(2.96\pm 1.49$ $\mathrm{s})$ decreased only slightly compared to No-eHMI, pedestrians recognized AV's non-yielding intention earlier $(\text{Q-Int = }7.35\pm 1.50)$, and perceived safety was markedly improved $(\text{Q-Safe = }6.84\pm 1.68$$)$. However, pedestrians' attention to the traffic environment decreased $(\text{Q-Env = }5.97\pm 1.99$$)$, and GC $(2.00\pm 0.66$$)$ increased. This suggested that Normal-eHMI based on a fixed activation distance increased pedestrian visual load, thereby distracting pedestrians and interfering with their overall judgment of the traffic environment.

The \textbf{IR-eHMI} condition achieved the shortest average SIT $(2.17 \pm 1.12$ $\mathrm{s})$, with GC $(1.67 \pm 0.62)$ at intermediate levels. Since Q-Int $(7.67 \pm1.24)$, Q-Safe $(7.33 \pm0.91)$, Q-Env $(7.17 \pm1.50)$ and Q-eHMI $(8.44 \pm1.38)$ were the highest, IR-eHMI is more effective for pedestrian decision-making. It also suggested that IR-eHMI helps pedestrians recognize AV's non-yielding intention earlier and make decisions more quickly without significantly increasing visual load compared to Normal-eHMI $(p<0.05)$. During the interaction, pedestrians perceived a high level of safety and maintained a high level of awareness of the surrounding traffic environment.

\begin{figure}[htbp]
   \centering
  \includegraphics[width=0.95\linewidth]{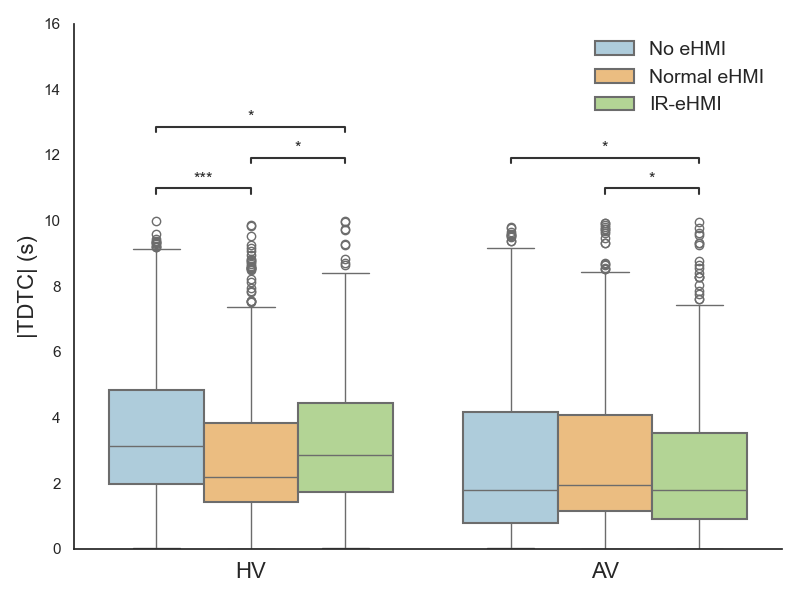}
  \caption{Comparison of $\vert$TDTC$\vert$ in S2 (*$p < 0.05$, **$p < 0.01$, ***$p < 0.001$).}
  \label{fig_result_case2}
\end{figure}

$\vert$TDTC$\vert$ serves as a key metric for evaluating pedestrian safety and in AV--pedestrian interactions \cite{TDTC}. Lower $\vert$TDTC$\vert$ indicates a higher potential risk. Fig. \ref{fig_result_case2} presents the distribution of $\vert$TDTC$\vert$ under different eHMI types for both the background vehicle (HV) and the interactive vehicle (AV) with pedestrians in the S2. 

From the perspective of HV safety, $\vert$TDTC$\vert$ across the three scenarios shew significant differences. The distribution of $\vert$TDTC$\vert$ under the \textbf{No-eHMI} condition is mainly concentrated in the $2$--$5$\,s range, with half of the values falling within this interval. Under the \textbf{Normal-eHMI} condition, $\vert$TDTC$\vert$ values became generally lower, with an average of $2.86$ $\mathrm{s}$ and half of the values concentrated between $1.4$--$3.8$ $\mathrm{s}$ indicating a potential increase in interaction risk. In contrast, \textbf{IR-eHMI} shew a clear safety advantage: the mean $\vert$TDTC$\vert$ increased to $3.25$ $\mathrm{s}$, and half of the values fell between $1.7$--$4.5$ $\mathrm{s}$, reflecting a more stable and safer interaction process with the background vehicle.

From the perspective of interaction safety with the AV, significant differences were observed only between \textbf{Normal-eHMI} and\textbf{ IR-eHMI} conditions $(p=0.035)$. Under the \textbf{No-eHMI} condition, $\vert$TDTC$\vert$ exhibited the widest dispersion, with a mean of $2.85$ $\mathrm{s}$ and half of observations falling within $0.77$–$4.18$ $\mathrm{s}$, indicating relatively high uncertainty in pedestrians’ responses. Under the \textbf{Normal-eHMI} condition, the mean $\vert$TDTC$\vert$ was relatively large and appears safer in terms of overall distribution. The average $\vert$TDTC$\vert$ under the \textbf{IR-eHMI} condition was slightly lower compared to \textbf{Normal-eHMI}. This was due to the reduced number of prompts. Nevertheless, in contrast to the \textbf{No-eHMI} condition $(\text{First Quartile}:0.77$ $\mathrm{s})$, the first quartile of the $\vert$TDTC$\vert$ distribution with \textbf{IR-eHMI} $(\text{First Quartile}:0.92$ $\mathrm{s})$ shifted upward, for the majority of trials.

In summary, the results shew pedestrians could infer the AV’s intention more easily through implicit interaction in the AV non-yielding scenario than in the AV yielding scenario. Although the Normal-eHMI provided an effective warning function that helps pedestrians interpret the AV’s intent more efficiently, it attracted excessive attention and led to distraction in some cases. By contrast, IR-eHMI reduced the number of eHMI prompts by $40\%$ and promoted more consistent pedestrian crossing decisions without imposing additional visual load, thereby maintaining overall interaction safety in multi-object interaction scenarios.

\subsubsection{In a case, IR-eHMI reduced distraction} 

\begin{figure}[htbp]
  \centering
  \includegraphics[width=0.95\linewidth]{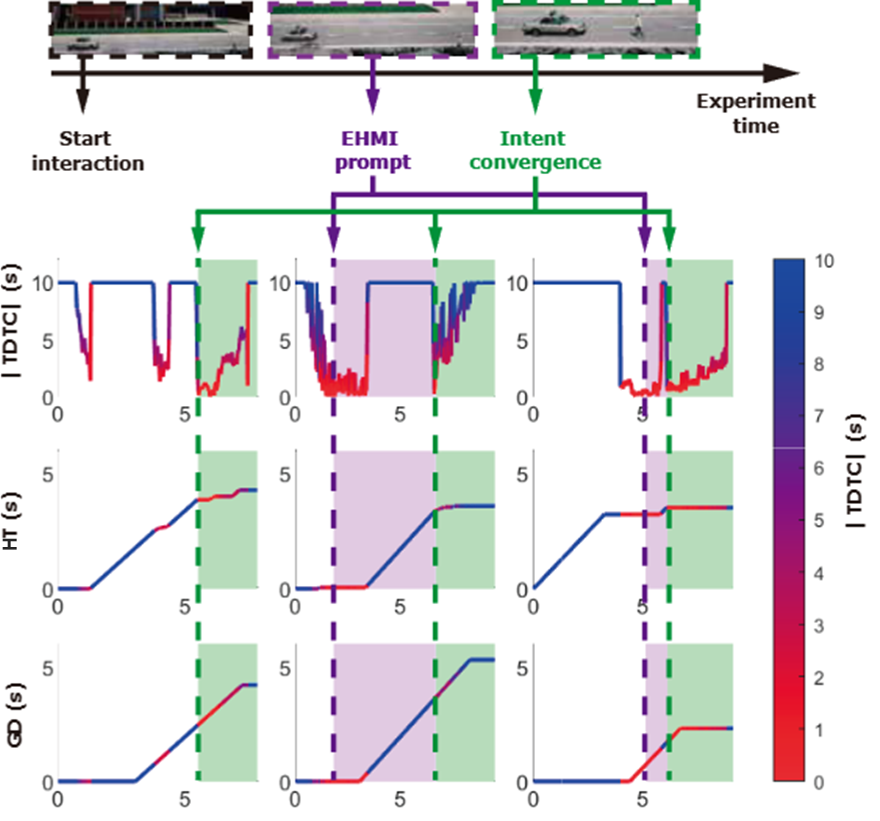}
  \caption{Comparison of critical indicators in a case}
  \label{fig_result2}
\end{figure}

In this case study, the behavioral characteristics of a participant in S1 under three different eHMI modes and compared, as illustrated in Fig. \ref{fig_result2}. 

Under the condition of \textbf{No-eHMI}, pedestrians started to cross the street after looking at the vehicle for $2.46$ $\mathrm{s}$, and $\vert$TDTC$\vert$ dropped rapidly during this period, indicating that pedestrians and vehicles reached an implicit consensus on intention. The total GD was $4.19$ $\mathrm{s}$, and the GC was $3$ during the crossing process, indicating that they were highly dependent on the vehicle dynamics for judgment. Under the condition of \textbf{Normal-eHMI}, the eHMI prompt was earlier than the pedestrian's initial gaze. After seeing the eHMI, the pedestrian paused and ultimately decided to cross the road after maintaining GD for $3.75$ $\mathrm{s}$. The changes in $\vert$TDTC$\vert$ were relatively stable, ensuring safety, but causing loss of efficiency and visual load.

Under the condition of \textbf{IR-eHMI}, pedestrians started to cross the street only $1$ $\mathrm{s}$ after receiving the eHMI prompt, with the GD of $2.31$ $\mathrm{s}$. During this process, $\vert$TDTC$\vert$ gradually decreased from a high value, indicating that the pedestrian made crossing decisions earlier and was able to quickly accept and effectively respond to IR-eHMI prompts. These results show that earlier eHMI prompts don't always improve performance. Adaptive eHMI based on intent consistency demonstrates superior effectiveness. 

In summary, several key findings emerged from this VR experiment:
\begin{itemize} 
\item Explicit interaction modes, including Normal-eHMI and IR-eHMI, generally improved safety and efficiency compared to implicit interactions, across AV yielding and AV non-yielding scenarios.
\item Normal-eHMII, which relies on a fixed activation distance, introduced substantial inter-individual variability and increased visual demand across AV yielding and AV non-yielding scenarios, thereby limiting its overall effectiveness. 
\item By selectively and adaptively activating eHMI cues based on the proposed IR algorithm, IR-eHMI effectively reduced gaze distraction while improving interaction efficiency and safety, demonstrating the potential of adaptive eHMI designs to enhance AV--pedestrian interactions. 
\end{itemize}

\section{Conclusion}

This study proposes an adaptive explicit interaction mode for AV--pedestrian communication through incorporating IR, termed IR-eHMI. The IR-eHMI monitors the dynamic evolution of traffic states and integrates these with pre-calibrated parameters to infer real-time cooperation states. By identifying intent consistency and potential conflicts, IR-eHMI enables timely and adaptive activation of explicit interaction cues. The proposed mechanism is implemented and evaluated on a VR-based platform supporting PIL experiments. Two distinct scenarios are considered: S1, in which the AV yields, and S2, in which the AV does not yield. The results corroborate prior concerns that conventional eHMIs may induce distraction and limit overall effectiveness. In contrast, IR-eHMI significantly enhances pedestrian crossing decision efficiency ($+12.8\%$ in S1 and $+13.0\%$ in S2), reduces gaze distraction ($-17.1\%$ in S1 and a $40\%$ reduction of eHMI prompts in S2), and maintains interaction safety, as evidenced by both behavioral metrics and questionnaire responses. Overall, the proposed mechanism introduces a new procedural paradigm for AV--pedestrian cooperation, and develops an effective evaluation framework, offering a promising direction for developing more intelligent and context-aware eHMI systems adaptable to diverse traffic scenarios.


\newpage
\section{Biography}
\begin{IEEEbiography}[{\includegraphics[width=1in,height=1.25in,clip,keepaspectratio]{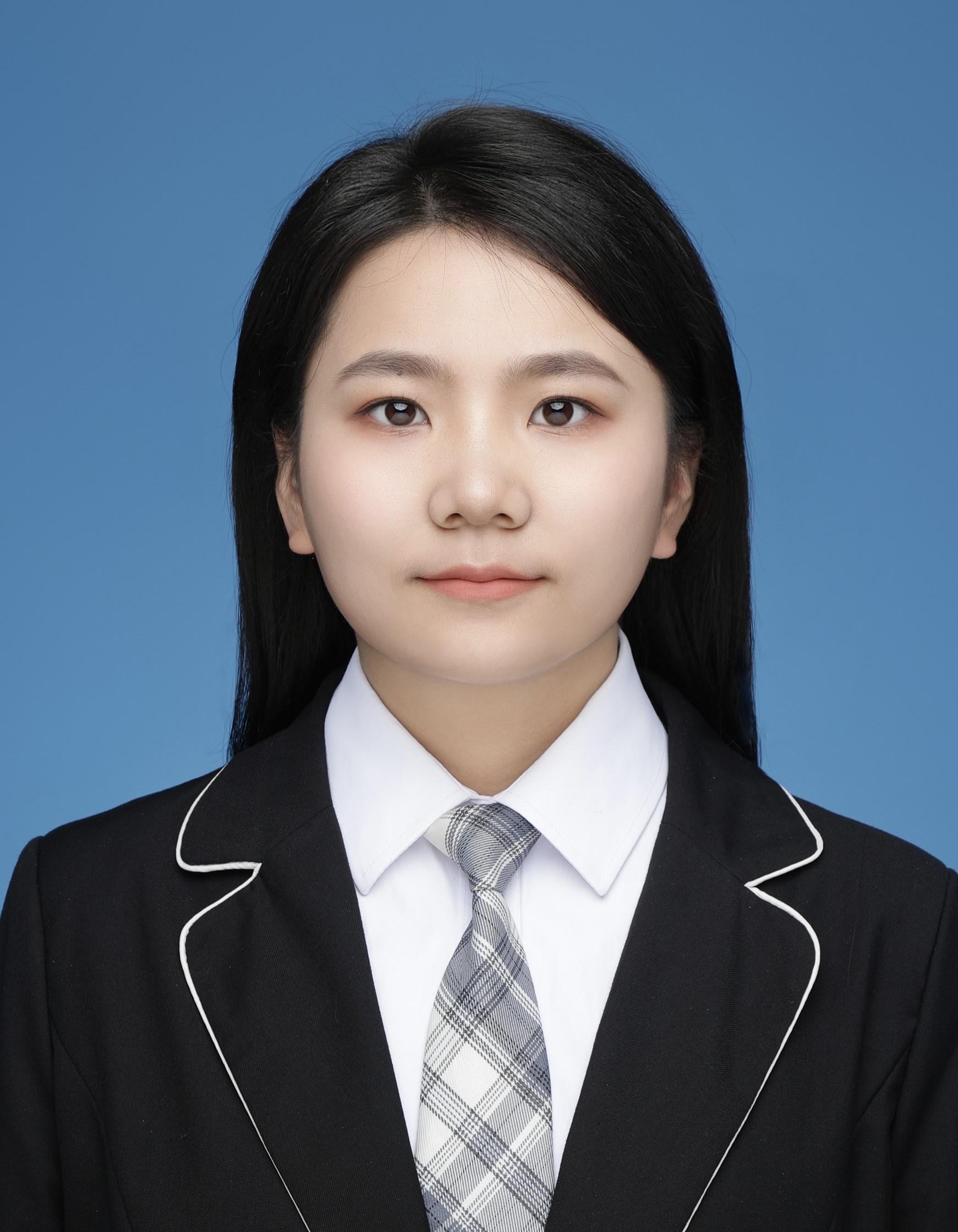}}]{Boya Sun} is currently a Ph.D. candidate in the College of Transportation at Tongji University. She received her B.S. degree from Southwest Jiaotong University. Her primary research interests focus on autonomous vehicle–-pedestrian interactions, as well as the evaluation and optimization of external human-–machine interfaces.
\end{IEEEbiography}
\begin{IEEEbiography}[{\includegraphics[width=1in,height=1.25in,clip,keepaspectratio]{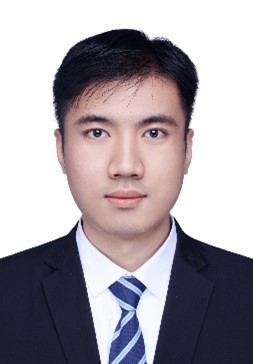}}]{Haotian Shi} received dual B.S. degrees in Computer Science and Energy and Power Engineering from Tianjin University, Tianjin, China, in 2017, an M.S. degree in Power Machinery and Engineering from Tianjin University in 2020, and M.S. and Ph.D. degrees in Transportation Engineering and Computer Sciences from the University of Wisconsin–Madison, Madison, WI, USA, in 2022 and 2023, respectively. He is currently a Tenured Associate Professor with the College of Transportation Engineering, Tongji University, Shanghai, China. His current research interests include autonomous driving behavior modeling and control optimization, intelligent connected vehicle testing and evaluation, and AI-empowered autonomous transportation systems..
\end{IEEEbiography}
\begin{IEEEbiography}[{\includegraphics[width=1in,height=1.25in,clip,keepaspectratio]{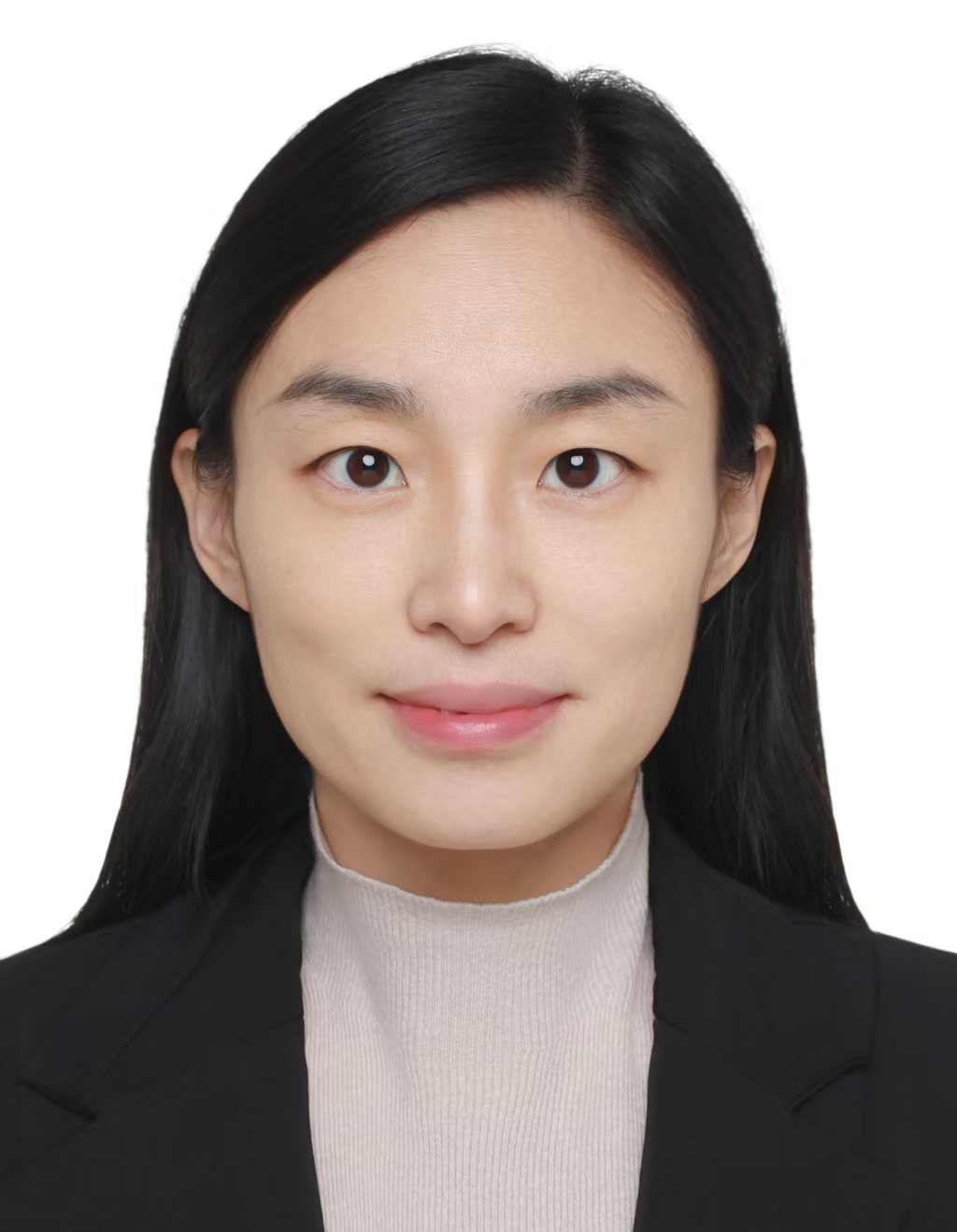}}]{Ying Ni} received the B.E. degree from the School of Southeast University, China, in 2005, and the Ph.D. degree from the Technical University of Darmstadt, Germany, in 2009. She was a Visiting Researcher with the University of California, Berkeley, USA, in 2014. She is currently an Associate Professor with the School of Transportation, Tongji University. Her research interests include modeling and simulation of interaction behavior under mixed traffic flow.
\end{IEEEbiography}
\begin{IEEEbiography}[{\includegraphics[width=1in,height=1.25in,clip,keepaspectratio]{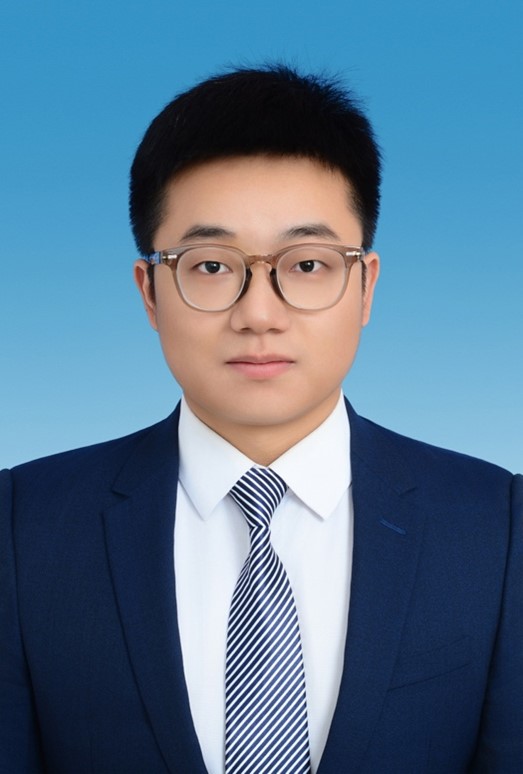}}]{Shaocheng Jia} is currently a Research Fellow at National University of Singapore. He received his B.Eng. degree from the Department of Electronic Information Engineering, China University of Petroleum (Beijing), M.Eng. degree from the Department of Automation, Tsinghua University, and Ph.D. degree from the Department of Civil Engineering, The University of Hong Kong, in 2018, 2021, and 2025, respectively. His research interests include connected and automated transportation, intelligent perception and control, stochastic modeling and optimization, and artificial intelligence. He has published over 20 papers in prestigious journals and conferences, including Transportation Science, TR-Part B, TR-Part C, IEEE TITS, ISTTT, and IEEE ITSC.
\end{IEEEbiography}
\begin{IEEEbiography}[{\includegraphics[width=1in,height=1.25in,clip,keepaspectratio]{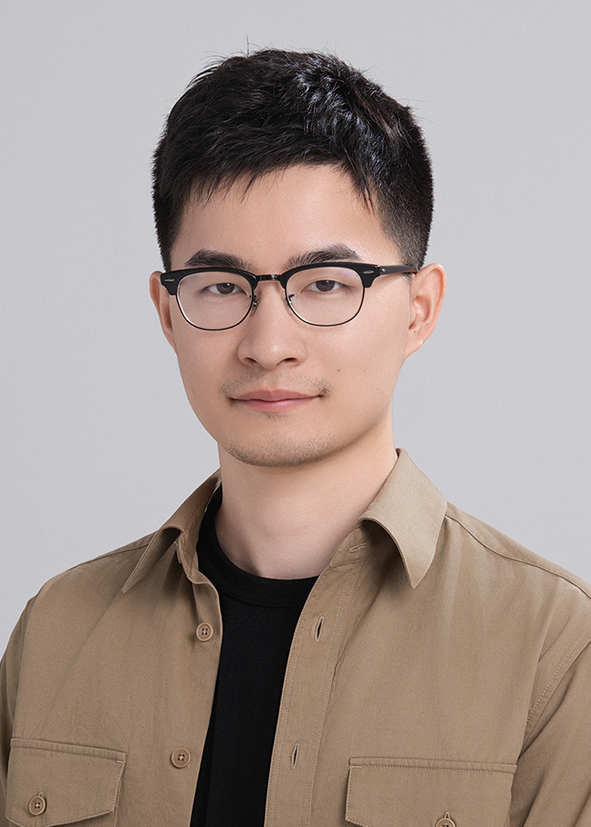}}]{Haoyang Liang} is currently a Postdoctoral researcher at Tongji University and a member of IEEE. He received his B.S. degree from the Department of Civil Engineering, Tsinghua University, and the Ph.D. degree from the University of Hong Kong, in 2018 and 2022, respectively. His research interests include crowd dynamics, pedestrian--vehicle interaction, virtual reality, and human factors. He has served as a workshop organizer at IEEE ITSC and a reviewer for journals including IEEE TITS, IEEE IoT, TBS, etc. Since 2023, he has been the principal investigator of TransCAVE LAB at Tongji University, focusing on developing digital twin platforms for mixed traffic simulation and testing.
\end{IEEEbiography}
\vfill

\end{document}